# Attitude, Aptitude, and Amplitude ( AAA ): A framework for design driven innovation

**Mohammad Lataifeh**
*University of Sharjah, UAE.*

### Abstract

*If creativity is the heart and soul of innovation, design is the main catalyst of this process. To design in a very simple sense, is to move from a certain situation into a better one; this amelioration can manifest in new products, services, processes, or merely by exposing the real problems at hand. Settling on a high momentum as an immersive human-centered method, design thinking has been promoted by many scholars and academic institutions for its strategic potential in solving different contemporary challenges faced by many organizations in different domains. This paper present a practical framework for how this form of intelligence can be introduced, learned and disseminated across the organization for a sustainable innovation to be achieved.*

### Introduction

Contemporary organizations are faced daily with new breeds of challenges fuelled by turbulent markets, advanced technology, globalization, environmental, and social policies. Although not as entangled as they are today, these challenges have evolved within different domains and scopes since the dawn of philosophy. Their phenomenal manifestations in modern history were tagged using different identities like ill-[structured , defined, or posed] problems (Mason & Mitroff, 1973; Newell, 1969; Reitman, 1964),  messes (Ackoff, 1979; Horn, 2001), and wicked problems (Rittel & Webber, 1973; Simon, 1973).

To maintain a competitive edge—even to remain in the market—leading through innovation strategy has been adopted by many organizations. In fact, many countries are embedding the topic within curriculums and educational systems from elementary to graduate schools (Beckman & Barry, 2007).

Being highly interconnected, volatile, and embedded within sociotechnical situations, contemporary challenges are invariably crossing the disciplinary boundaries to defy a structured, or algorithmic problem solving methods. To be able to innovate, an organization is engaged with a constant re-evaluation for the situation materials (context, players, and actions), a process that is crucially needed to define problem space before revealing the new permissible means of a resolution. This is a complete shift from the irresistible assumptions that the cause of a problem is "*out there*" rather than "*in here*" (Meadows, 2008, p. 4), within the specific groups, their interactions, and the processes followed to manage their wickedness.

### Design Thinking

Design has found it is way to the agenda of many organizations. From being the "thing" in business school (Garrett, 2014), innovation driver in public sectors (Bason et al., 2013), to an overall strategy for organizations (Mootee, 2013) and governments at large (Clinton Global Initiative, 2012).

Notwithstanding the nuances between an academic (Buchanan, 1992; Cross, 1982, 2001; Krippendorff, 2006; Lawson, 1980, 2005; Rittel & Webber, 1973; Schön, 1983; Simon, 1969), and a practical account of design thinking (Boland, Collopy, Lyytinen, & Yoo, 2008; Boland & Collopy, 2004; Brown, 2009; Dunne & Martin, 2006; Martin, 2008, 2009, 2011),  design thinking can be simply defined as 'how designers think'. Their logic, activities, processes, tools and actions that form their distinct profession.

 Design thinking is proposed here as a form of intelligence that can go far beyond merely problem solving. When compared to the latter, design cognitive activities "*recruits a more extensive network of brain areas*" while working to evaluate and modulate appropriate responses for uncertain conditions (Cross, 2010, p. 103). Designer's cognitive activities in framing, conceptualizing, visualization, reasoning and discovering are iteratively and creatively moving between solution and problem spaces in an orderly, purposefully, and intentionally manner that demonstrates different mental and physical activities. Hence, the term is used as a process, a communicator, an eco-system (Botero, Kommonen, & Marttila, 2010), even a paradigm (Dorst, 2011) to cultivate a fruitful inquiry within the organization, shifting design focus from objects to objectives.

### Why Design Thinking?

As an immersive human-centred method, design thinking strategic potential has been established within business and management studies (Boland & Collopy, 2004; Brown, 2008; Dorst, 2011; Gibson & Brown, 2009; Leavy, 2011; Martin, 2009; Romme, 2003; Verganti, 2009), development and planning of Information Systems (Du, Jing, & Liu, 2011; Luebbe, Weske, Edelman, Steinert, & Leifer, 2010; Plattner, Meinel, & Leifer, 2010), leadership (Norton, 2012), biotechnology (Friedman, 2011), military (Bullock & Vitor, 2010), nursing (MacFadyen, 2014), even literacy (Purdy, 2014) among many other areas.

Design thinking adoption to drive innovation has indeed pushed to reform design  educational systems (Norman & Klemmer, 2014) and its relation to other disciplines,  some of which, may have already started as per the examples of Hasso-Plattner-Institute in Germany, and the Design School (d.School) at Stanford University, where students from all disciplines get together to study, work and solve problems based on this approach. Design thinking in such context takes wider perspectives from the empathetic transformation of needs into opportunities (Brown, 2009), to the focus of innovations in systems, organizations and networks at large (Ansell & Torfin, 2014; Davis, 2010; Du et al., 2011).

With an emphasis on the desired state of affairs (Simon, 1969, 1973), the examination of an issue through the design thinking lens seems to help businesses overcome some of the limitations of traditional disciplinary approaches within organizations, particularly where prescriptive routines, or narrow disciplinary boundaries prohibit exploration of new ways to view or manage these issues.

### *Operationalization of Design Thinking*

While design thinking can be expressed as a ubiquitous human activity (Razzouk & Shute, 2012) that does not require immense training or a professional label to experience, it is important to submit that benefit realizations differ according to *designers'* level of expertise established in the literature (Candy & Edmonds, 2006; Dreyfus & Dreyfus, 1980, 2005; Lawson & Dorst, 2009) creating unique sets of characteristics that distinguish these levels among individuals. Collectively they describe the overall *designerly* process, particularly the exhibited mindsets that could be embodied by others in attempting to address a particular issue in a new domain.

Despite the acknowledged differences in levels of expertise, educators and professionals within the business and management world are increasingly mobilizing design in their work. Kimbell (2011, 2012) offered a pair of concepts to consider as a basis for re-thinking design thinking 'design as a practice', and 'design in practice', both of which are important to break from the duality between design and designers, and in a way mitigate possible influence of levels of expertise as a barrier for the embodiment and utilization of the process, since building a creative, agile, and innovative organizational culture starts with engaging constituents within; equipping them with skillset that complements their own vertical expertise to lead the process instead of relying on external expertise to do so. A novel pragmatic framework is discussed next.

### *AAA Design Transformational Framework*

It is important to highlight that design activities and the mindsets experienced during the design process necessarily demand a particular cultural and organizational setting including team compositions and expertise, design space, management style, and individual drive to name a few. Therefore, the proposed process within the AAA framework should be placed within a larger intervention framework that addresses these settings (depicted below in Figure 8), discussed further in a separate work (Lataifeh, 2015).

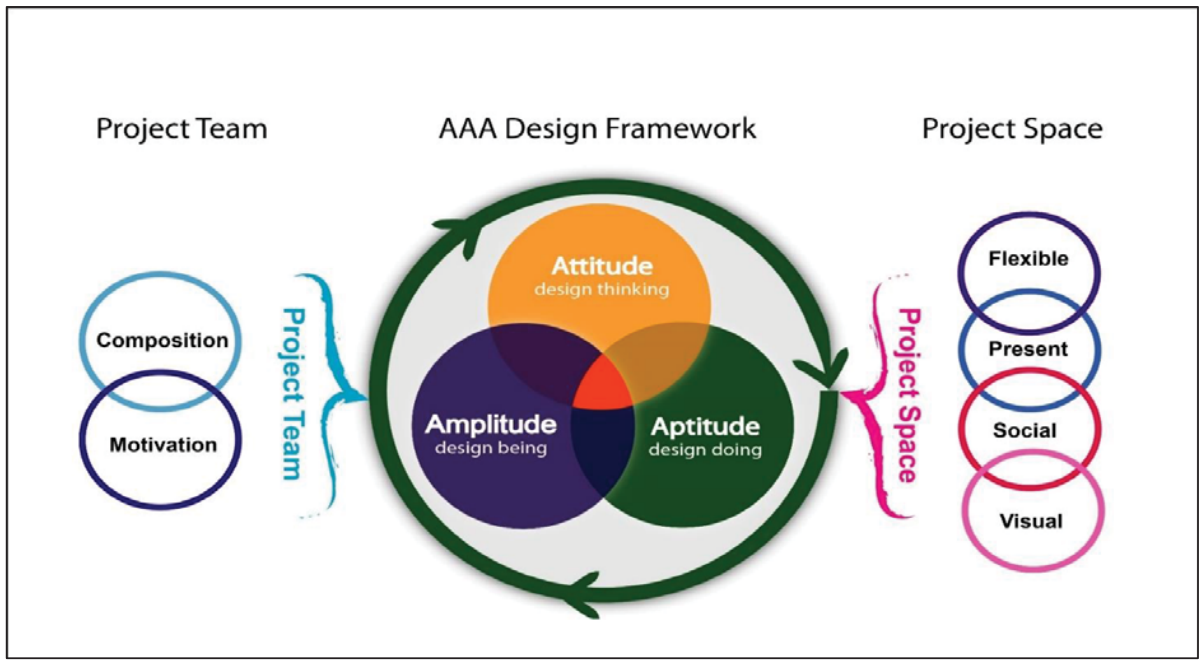

**Figure 8- Design Intervention Framework (Lataifeh, 2015).**

Nonetheless, the AAA framework (Figure 9) simplifies a pragmatic design transformational process that goes through three different stages: attitudes (design thinking), aptitudes (design doing), and amplitudes (design being). The following sections in this paper discuss these three stages in terms of how they can be approached, their boundaries, objectives, materials, integration, and their value to the overall process of institutionalizing a design driven innovation culture.

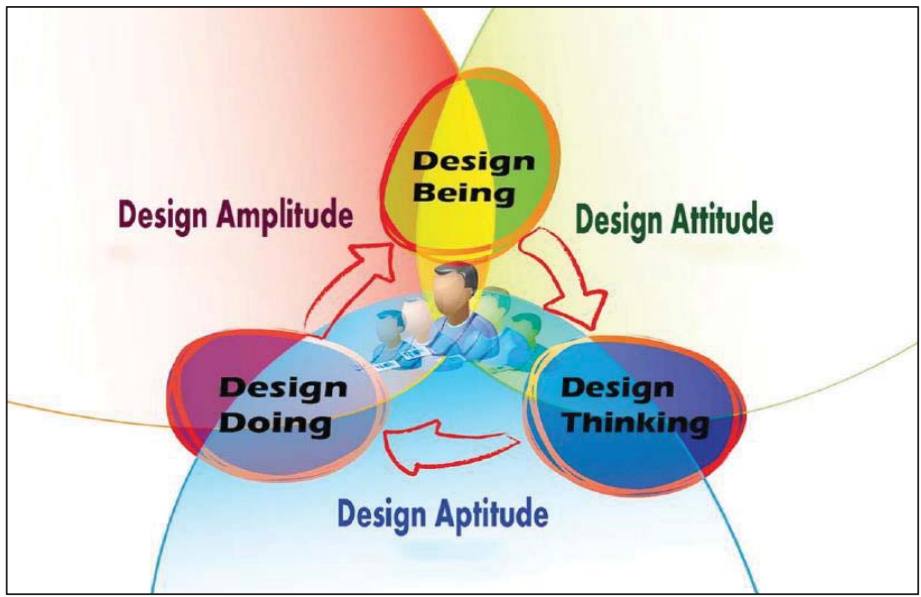

**Figure 9- AAA design framework (Lataifeh, 2015)**

*Attitudes (design thinking)*

The first and probably the toughest challenge for design and design thinking to address is to claim a seat at the table [to be accepted as a legitimate business tool] (Holloway, 2007; Repisky, 2014), and to be seriously considered beyond the traditional boundaries of products and services. To be positioned as a form of intelligence (Cross, 2010), guiding transformational results for the organization and society at large (Pastor, 2013) it all starts with a basic, but a profound notion of attitude, a *"little thing that makes a big difference"* (Ballon & Skinner, 2008, p. 218).

Attitudes have been the single most researched topic in social psychology, but the term is often left vague in the literature (Augoustinos, Walker, & Donaghue, 2006, p. 113). Attitudes are typically defined as "*predispositions to respond in a particular way toward a specified class of objects*" (Rosenberg & Hovland, 1960, p. 1). In order to make explicit the components of attitudes, Triandis (1971, p. 2) defined attitudes as *"an idea charged with emotions, which predisposes a class of action to a particular class of social situations"*. An attitude is therefore composed of three different and consistent components or responses:

    **A.** A cognitive component or an idea that is part of some category used by humans while thinking (food, cars, sickness, etc.)

    **B.** An affective component, that is, the emotion which charges the idea (positive or negative)

    **C.** A behavioural component, as in a predisposition to act (explicit or implicit)

(Rosenberg & Hovland, 1960, p. 2; Triandis, 1971, p. 3)

The consistencies of responses (thinking, feeling, and acting) of an individual towards a certain situation (stimuli), represent an individual attitude which is developed in order to *understand* and *adjust* to a complex world, to *protect self-esteem*, and to *express fundamental values* (Triandis, 1971, p. 101). Attitudes can coexist implicitly and explicitly (Wilson, Lindsey, & Schooler, 2000), as they are similar to personality traits being latent and hypothetical constructs. Inaccessible to direct observations, attitudes differ by being inferred from measurable responses of the attitude components seen in cognitive, affective, and conative (behavioural inclinations and intentions) responses (Ajzen, 2005, pp. 3–6).

One of the ambiguous issues for researchers in this domain is the relationship between attitudes and behaviour (Augoustinos et al., 2006, p. 25). Since attitudes are inferred from behaviours, a direct link may be readily conceived to predict the latter. In reality though, a simple straightforward link cannot be easily construed (Fishbein & Ajzen, 1974; Triandis, 1971). People behaviours may contradict their beliefs, or indifference to their attitudes, which empirically undermines the validity of a consistency of behaviours that reflects the presumed attitude or predisposition (Ajzen, 2005, p. 33). Which nonetheless, confirms that "*behaviours can cause attitudes as much as the other way around*" (Augoustinos et al., 2006, p. 25).

In Addition to variable individual moderators like self-monitoring, self-consciousness, or self-awareness (Ajzen & Fishbein, 2005), behaviours are contextually determined by more than attitudes including norms, habits, expectations and reinforcement (Triandis, 1971, p. 25). Situational variables not only impact behaviours independently of its stable dispositions, but they "*can also moderate the effects of attitudes or personality traits*" (Ajzen, 2005, p. 41).

Intentions on the other hand, are found to forge an accurate predictor for a variety of actions, particularly when people have control over performance like skills, emotion, and the opportunity to act upon intentions (Ajzen, 2005, pp. 99–115). Hence, intentions become central to the attitude-behaviour relationship, and the base for the theory of planned behaviour (Figure 10); where attitudes, norms, and perceived control over behaviour could predict intentions that can account for a large proportion of variance in a behaviour (Ajzen, 1991, 2005).

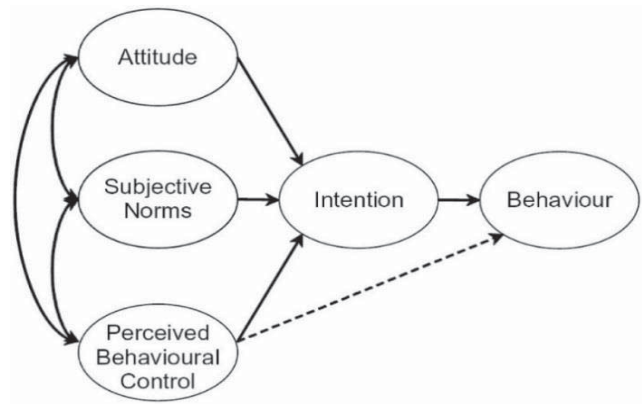

**Figure 10- Theory of planned behaviour (Ajzen, 2005)**

Attitudes can be learned and relearned (changed), either by new information received (particularly family, peers, friends or other media) that changes the *cognitive component* of an attitude, or by a direct experience with the object of the attitude through the *affective component* (Triandis, 1971, pp. 142–146). The latter is found more effective than secondary information (Fazio & Williams, 1986; Sherman & Fazio, 1983).

In terms of communicating new information on changing behaviours, McGuir (1968a, 1968b) proposed a process with specific stages that information needs to go through to have an observable effect: attention, comprehension, yielding, retention, and action. Each stage has its own variables that define the level of its achievement to proceed to the next stage. For instance, attention may be reduced by distractions; comprehension and yielding are influenced by the receiver level of intelligence; retention is influenced by message intensity or duration, as well as interference of other messages. As for actions, there could be a million reasons why it may not take place. It suffices to say that the extent to which attitudes can be changed depends upon multiple factors related to the situation, the group, the individual, the message, the source of the message, the medium, and the presentation as elaborated by Triandis (1971, pp. 142–200).

### Aptitudes (design doing)

Having influenced a positive attitude, or at least gained participants' attention, the aptitudes stage is concerned with the culmination of participants' creative confidence (Kelley, 2012), to trust their own creative skills. Creative confidence is seen as the pinnacle of this learning approach as deployed by design thinking schools in Potsdam and Stanford (Jobst, Ko, Lindberg, Moritz, & Meinel, 2012; Rauth, Köppen, Jobst, & Meinel, 2010). Equipping participants with different design tools deliberately selected from a wider toolset (Dubberly, 2008; Hanington & Martin, 2012; Kumar, 2013) to fit different implementation/applications, participants gain experience going through the processes of design within a contextualized environment. In doing so, creating the opportunity for a direct experience with the cognitive element as a favourable approach to affect behaviours (Fazio & Williams, 1986; Sherman & Fazio, 1983), which aligns with the theory of planned action (Ajzen, 2005) in enhancing control over performance.

Starting with divergent thinking tools to explore the problem space, and closing with convergent tools to establish team consensus on solutions, participants are challenged to use different mindsets—epistemological stances or world views—utilizing different thinking modes as they embody the logic of design (Burnette, 2009). The process attempts to transform a static mindset into an evolving stage of *mindshifts* (Goldman et al., 2012, p. 15), as a continuous transition of a participant's instincts and orientations to come into the new being of design thinkers.

To experience the impact of design thinking, participants have to complete a full design journey, from uncovering problems to devising solutions. Therefore, a design thinking process is needed to be followed here. Dubberly (2008) offered a compendium of different design processes or models used in the industry, grouped according to their flows, objectives and context. The majority of these models located different stages between problem and solution space (Owen, 2007) as they coevolve together (Dorst & Cross, 2001). Situated in particular contexts, the nature of the problem solving process itself shapes the solution (Rowe, 1987), with the problem definition and solution becoming mutually intertwined (Zdrahal, 2007). The problem cannot be clarified completely without choosing a solution during the problem solving process, and the method for solving it, cannot be selected without understanding the problem. Any particular formulation of the problem, hints at a different solution. Therefore, the spaces between problem definition, synthesis and evaluation are inseparable as they all occur at the same time. Solutions are often masked by the common ill-defined nature of these (design) problems.

Researchers have produced several process models during the last decade, from Lawson (2005, p. 34,149), Cross (2000, p. 36), Design Council (2005), and Kumar (2009), to the most recent models from the HPI d.school[8] (2010), and Liedtka & Ogilvie (2011), both of which are depicted in    Figure 11.

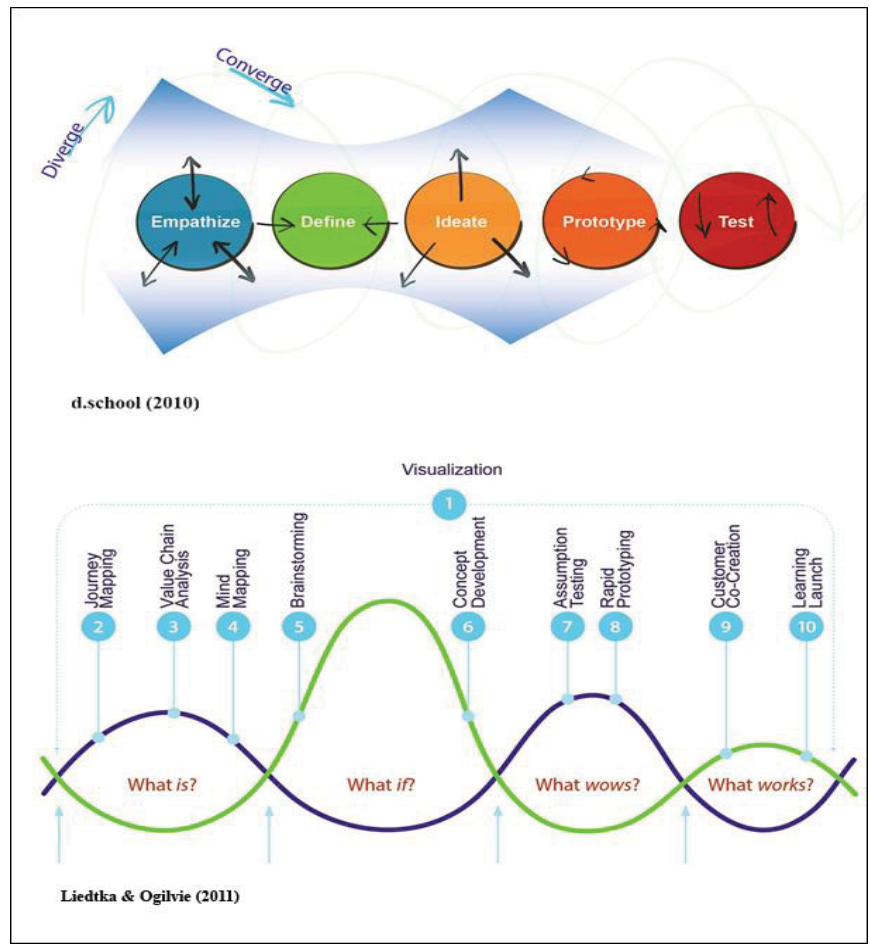

**Figure 11- Most recent design thinking processes**

---

[8] Design Schools jointly managed by IDEO and Hasso Plattner Institute at Potsdam and Stanford.

The five stages of the design process are now described including a summary of the mindsets and the tools proposed for each stage. It is important to note that participants may move iteratively between these stages to satisfy a particular need, rather than follow a linear one directional approach.

### Empathize

The empathy concept has been used from early twentieth century as a translation for an older German concept of *Einfühlung,* which encompasses notions such as sympathy, understanding, and role/perspective taking, while framing or tapping into other people's feelings and behaviours within psychological tradition (Nowak, 2011; Wispe, 1990, pp. 17–19). The objective of empathy is to tune into others' *wavelength* to understand with them, not about them (Tudor, 2011, p. 41). Sensing context and knowing people entails high degrees of sensitivity in observing, dialectical understanding, and temporarily *living the life* of others (Rogers, 1995, p. 142), within a process that includes three different components: *1) an affective response to share other's emotional state, 2) a cognitive capacity to experience other's perspectives, and 3) a monitoring mechanism to track the origins of experienced feelings (self Vs. others)* (Lamm, Batson, & Decety, 2007, p. 42).

Empathy is therefore a central piece for an affective engagement and team connectedness, as it enhances group coherence beyond self-boundaries (Pavlovich & Krahnke, 2011), allowing participants to develop a collaborative understanding for the real issues at hand.  Furthermore, if empathy is the core of human-centred design, ethnographic tools are instrumental for the materialization of this process, particularly observation of users' interactions with each other or with their material landscape, informal conversation with deep listening to elicit mutual trust and tolerance, and role playing/shadowing that enables living the world as experienced by others (McDonagh, 2010).  As a result, the mindsets of openness, curiosity, inquisitive, failure permitting, patience, and non-judgmental interactions are essential for the participants to experience empathy.

Some of the design tools that can be used in this stage include but are not limited to camera study, empathy mapping, journey mapping, behavioural mapping, stakeholder mapping, social network mapping, personal inventory/cultural artefacts, users shadowing and immersion in day in the life of a user.

### Define

Taking the scattered findings forward, participants reflectively unpack or synthesize meaningful insights into problem framing (Cross, 2003; Schön, 1983). The problem space is not given but rather constructed by participants (Schön, 1991), and continues to be reframed (Choulier, 2011) during this *reflection-in-action* process, charged with setting problem boundaries to *"select particular things and relations for attention, and impose on the situation a coherence that guides subsequent moves"* (Schön, 1988; as cited in Cross, 2006, p. 80), while drafting possible solutions to be evaluated (Dorst, 1995). Framing problems is fundamental for a successful design process (Cross, 2011), and a key feature of a developed design expertise (Cross, 2004).

Problem framing is rarely completed in a single stretch at the beginning of a design process (Schön, 1988).  As an evolving process, every frame constructed leads to a conjectured solution, hence having multiple accepted/satisfying solutions, rather than one correct solution is a known feature of the design process (Visser, 2009). The alternatives serve as *means of problem-analysis* (Cross, 2006, p. 17). While problem identification is a challenging task (Pacanowsky, 1995), these conjectures mitigate the risk of falling too soon on a fixation of a narrowly framed problem (Liedtka & Ogilvie, 2011, p. 24), through a reflective approach that entails the willingness to endure suspense and mental unrest in the search process.  An approach that assists the team in overcoming the "*inertia that inclines one to accept suggestions* [ideas] *at their face value"*(Dewey, 1910, p. 13).

The main mindsets encouraged at this stage include visualization, finding patterns, exploring systems, analyzing/sorting insights, developing a point-of-view, drawing the big picture, sensing gaps, identifying opportunities and decision making. These mindsets can be experienced using a variety of design tools including but not limited to mind mapping, fishbone diagrams, extreme users, ERAF (entity, relations, attributes and flows) diagrams, Venn diagrams, analogous empathy, point-of-view analogy, hits and highlights, statement starters, reverse assumptions and Pareto analysis.

### *Ideate*

The ideation stage is where participants diverge widely on the developed insights, generating as many ideas as possible. As they begin the exploration of the solution space (Lindberg, Meinel, & Wagner, 2010), it is critical to encourage a state of *flow* (Csikszentmihalyi & Csikszentmihalyi, 1992, p. 30)  by focusing on the engagement process, to enable a fluent ideation that also demands an adjourned judgment. The ideation process is guided by the developed insights as objectives only, rather than an evaluation criterion that can hinder the creative process (Vangundy, 2005).  The inevitable uncertainty must be embraced to step away from familiar or known solutions (Liedtka, 2011).  Nonetheless, ideation should be structured according to the predefined insights in order to remain focused, and time-boxed as an activity with turn-away intervals, to allow for non-conscious processing or creative incubations to occur (Ellwood, Pallier, Snyder, & Gallate, 2009), since no one can be in constant *flow* all the time (Csikszentmihalyi, 2003, p. 71). The objective of this stage is to cultivate participants' explicit consensus, and to uncover tacit connections between entities, objects, domains, and users within the organization based on representatives' inputs.

The mindsets encouraged at this stage include visualization, humor, happiness, talkative, artistry, intuitive, playfulness, seeing the big picture and freely toying with ideas. These mindsets can be experienced using several tools including but not limited to powers of ten, brainstorming, brain writing, KJ technique, affinity diagrams, card sorting, free-listing, imposed constraints, laddering questions, idea boxes, SCAMPER (substitute, combine, adapt, modify, put to another use, eliminate and reverse), concept generating matrices, and concept mapping/sorting.

### Prototype

During the prototyping stage, ideas (internal mental images) are rapidly externalized into sketches, stories, scenarios, models and other artefacts. These low fidelity prototypes provide the vehicle for further experiments (Schön, 1983, p. 174), and the materialized representations become more accessible for direct interactions and conversations that move between criticism and discovery (Cross, 2011, pp. 12–13); helping to experience the shift from thinking to doing. Participants' reflective and experimental actions then lead this discovery process, assisted by the materials' talk-back (Schön, 1983, p. 280).

Prototypes provide a concrete realm as an escape from operating only within mental abstractions. The design process continues to *fluidly* move between the abstract and the concrete (Beckman & Barry, 2007, p. 50), an oscillation that is energized by low fidelity prototypes eliciting a rapid evaluation and feedback, to manage the uncertainty within complex systems landscapes (Gerber, 2009). Rapidly constructed prototypes must be good enough to learn, rather than exhaustively tested at this stage (Liedtka & Ogilvie, 2011, p. 33). In contrast, costly (time and efforts) high fidelity prototypes—of near finished systems or products—could hinder this fluency, as participants get distracted with the operational details and less with contemplating improvements (Atladottir, Hvannberg, & Gunnarsdottir, 2011).

Apart from their valued outcomes, low fidelity prototypes are also important as an *enactment* for the design practice, with a positive psychological effect on participants, as they reframe failure into learning opportunities while refining their solutions. A process that overall enhances confidence in participants' creative abilities (Gerber & Carroll, 2012, p. 70). Therefore, prototypes not only serve as a manifestation of ideas, but also as a stimulus and filtering mechanism to guide ideations in new directions, and assist in concisely defining the problem space (Lim, Stolterman, & Tenenberg, 2008), within a highly and seamlessly integrated manner.

The mindsets sought for this stage include experimentation, action-oriented, openness, playfulness, collaboration, creative use of raw materials, process oriented, evaluative thinking, contemplation, environment sensitivity and contextual thinking. The list of design tools that can assist here includes generative research, design charrettes, storyboards, storytelling, bodystorming, creative tool kits, wizard-of-oz, space prototyping, and AEIOU (activities, environment, interactions, objects and users).

### Test

Testing and prototyping are almost inseparable. The two stages are used in tandem, with a feedback loop for the devised solution. Testing is not only central to validating requirements fulfilments, but as a further understanding to the solution that leads to uncovering new requirements moulded by the preceding alternatives (Cross, 2011, pp. 16–17). This process can be linked to experiential learning theory (Kolb, 1984), where learning and knowledge creation occur within loops between active/synthetic, and reflective/analytic experimentations until a working combination (solution) is found (Beckman & Barry, 2007), as an emergence (Cross, 2006, p. 55) for the design enactment experienced within different design thinking stages.

The mindsets desired for this stage focus on context sensitive, tactical, risk taking, persistent, courage, and seeking support. As for the design tools that can be used here it may start with a variation of 2 x 2 matrices for contextualization, then move to PPCO (pluses, potentials, concerns and overcoming), storytelling, testing with end users, feedback capture grids, 'I like, I wish, what if', 'how-how diagrams', and stakeholder analysis.

### Amplitudes (design being)

Amplitudes as the third stage of the proposed design framework are constructed as a recursive process for the emerged knowledge. This process constitutes a ripple effect that transcends the implementing teams (initial design process participants), to the rest of the organization.

Implementing teams exposed to the design process experience, become the internal catalysts for design thinking with the attitudes, aptitudes, and local examples necessary to support the case for its value. Therefore, the message, source, channel, and presentation are all closely related, and shall be captured easily by constituent in demonstrating the reality of the situation.

Nonetheless, for the rest of the organization to act, this new knowledge must a) be materialized as a unique representation of a distributed understanding, and b) voiced by someone or some group (Taylor & Van Every, 2000, p. 243). After all, *"if cognition lies in the path of the action, then texts and conversations also lie in its path"* (Weick, 2009, p. 5). Text in this context includes all produced materials or artefacts as different tangible instances for the *objects of knowledge,* which remain projections-in-context, *simultaneously conceived as unfolding structures of absences,* rather than definitive things (Cetina, 2001, p. 191). Therefore, the mere existence of these materials is critical to support design conversations at an organization level, thus harnessing group knowledge, in order to sustain the transformational process.

### Conclusion

The design intervention framework (Figure 8) was proposed to use design thinking as main catalyst for innovating solutions for the contemporary sociotechnical challenges in large organizations. The framework is proposed as a pragmatic methodological approach that address project space and project teams variables to influence a positive environment for the AAA framework to be more effective.

The AAA design framework has been devised with different stages to introduce, enhance, and establish design competencies across the organization to seed the culture of innovation within. Although aligned with the prevailing design processes highlighted in previous research, the AAA framework was further developed to address the breadth of issues (intentions, cognition, abilities, etc.) to be practically approached in fulfilling the demand for cultivating innovative solutions from within an organization (inward sourcing) by its own constituent, instead of one size fits all solutions parachuted by external consultancies, not only without sufficient understanding to context, but with no empathy to constituents inside and outside these organizations.

To truly institutionalize innovation in large organizations, organizational management culture needs to be reinvented from the top down to empower local expertise, promote ownership over process, allow mistakes to happen, and reward ideas not just sales. All of these might be drastically different of how managers may have been trained to do their work, there will be impediments stemming from different organizational cults, and quite often, the risk of losing control might seem imminent, but a leap of faith is inevitable to attain a well worth success.